\documentclass{article}
\usepackage{arxiv}

\usepackage[T1]{fontenc}    
\usepackage{hyperref}       
\usepackage{url}            
\usepackage{booktabs}       
\usepackage{amsfonts}       
\usepackage{nicefrac}       
\usepackage{microtype}      
\usepackage{lipsum}

\usepackage{graphicx}          

\usepackage{float}
\usepackage{amsmath}
\usepackage{amsthm}
\usepackage{xcolor}
\usepackage{siunitx}
\usepackage{booktabs}
\usepackage{color}
\usepackage{caption}
\usepackage{subcaption}
\usepackage{amssymb}
\usepackage{graphicx}
\usepackage{picinpar}
\usepackage{url}
\usepackage{flushend}
\usepackage[latin1]{inputenc}
\usepackage{colortbl}
\usepackage{soul}
\usepackage{multirow}
\usepackage{pifont}
\usepackage{alltt}
\usepackage{enumerate}
\usepackage{siunitx}
\usepackage{epstopdf}
\usepackage{pbox}
\usepackage[vlined,linesnumberedhidden,ruled,resetcount]{algorithm2e}
\usepackage{tabulary}

\usepackage{graphics} 
\usepackage{graphicx} 
\usepackage{epsfig} 
\usepackage{enumitem} 

\usepackage{enumerate}
\usepackage{multirow}

\usepackage{amssymb}  

\usepackage{amsthm}

\newcommand{\nextnr}{\stepcounter{AlgoLine}\ShowLn}

\usepackage{array}
\usepackage{makecell}

\newtheorem{rem}{Remark}

\newtheorem{deff}{Definition}

\newtheorem{limm}{Limitation}

\title{Thermal Fault Detection and Localization Framework for Large Format Batteries
}

 \author{Sara~Sattarzadeh,
         Tanushree~Roy
         and~Satadru~Dey
\thanks{S. Sattarzadeh, T. Roy, and S. Dey are with the Department of Mechanical Engineering,
        The Pennsylvania State University, University Park, PA 16802, USA.
        {\tt\small \{sfs6216,tbr5281,skd5685\}@psu.edu}. This work is supported by National Science Foundation under Grants No. 1908560 and 2050315. The opinions, findings, and conclusions or recommendations expressed are those of the author(s) and do not necessarily reflect the views of the National Science Foundation.}%
 }

\begin{document}
\maketitle

\begin{abstract}
Safety against thermal failures is crucial in  battery systems. Real-time thermal diagnostics can be a key enabler of such safer batteries. Thermal fault diagnostics in large format pouch or prismatic cells pose additional challenges compared to cylindrical cells. These challenges arise from the fact that the temperature distribution in large format cells is at least two-dimensional in nature (along length and breadth) while such distribution can be reasonably approximated in one dimension (along radial direction) in cylindrical cells. This difference makes the placement of temperature sensor(s) non-trivial and the design of detection algorithm challenging. In this work, we address these issues by proposing a framework that (i) optimizes the sensor locations to improve detectability and isolability of thermal faults, and (ii) designs a filtering scheme for fault detection and localization based on a two dimensional thermal model. The proposed framework is illustrated by experimental and simulation studies on a commercial battery cell. 
\end{abstract}


\section{Introduction}
Safety is still one of the critical barriers for battery technologies. From battery control viewpoint, real-time diagnostics of battery faults is a key towards safer batteries. These internal faults can originate from wide range of factors such as manufacturing defects, abusive operating conditions, and internal degradation mechanisms induced by aging. Irrespective of the {physical cause} of the failure, many internal faults eventually manifest themselves as abnormal thermal behavior which may in turn lead to thermal runaway. Therefore, early detection of such thermal faults at their nascent stage is indispensable for battery safety. In this work, we specifically focus on thermal fault detection and localization in large format batteries such as prismatic and pouch batteries. 

\subsection{Motivation and challenges}
Thanks to the higher energy, higher power, longer life-time and longer operation-time, Lithium-ion batteries have become more popular in various applications. Safety against thermal anomalies is crucial in most of these applications. Thermal safety in Lithium-ion batteries is aided by a combination of installed temperature sensors and thermal management algorithms. In some applications such as passenger automotive and consumer electronics, typically one temperature sensor is deployed per cell due to cost effectiveness. However, some applications are highly safety critical in nature, such as electric mining vehicles, mining robots in coal mines, drones which move in explosive atmosphere \cite{dubaniewicz2013lithium,dubaniewicz2021thermal,faranda2019lithium, ma2021problems,caldwell2017hull}, wearable and implantable biomedical devices \cite{pu2015self,bock2012batteries}. Even with Explosion Proof (EX) enclosures in such applications, thermal runaway represents a critical risk \cite{dubaniewicz2013lithium,dubaniewicz2021thermal}. In these safety critical applications, it is reasonable to install multiple temperature sensors on a single cell as the safety risk outweighs the cost effectiveness. Motivated by this, we propose an algorithmic framework for thermal fault detection and localization in Lithium-ion batteries. This framework discusses multiple temperature sensors as well as single sensor scenarios accounting for a wide range of applications.

Compared to cylindrical cells, large format prismatic or pouch cells pose additional challenges in thermal fault diagnostics. Temperature distribution in cylindrical cells can be reasonably captured by one-dimensional distributed parameter model that accounts for distribution along radial direction while assuming uniform distribution along the cell height \cite{kim2014estimation}. On the other hand, notable variation in temperature has been observed along the length and breadth of large format pouch and prismatic cells which leads to two dimensional distributed parameter models \cite{Yazdanpour,Debert2013,tian20173,sattarzadeh2021real}. Such variation makes the fault detection and localization in this two-dimensional space a non-trivial issue. Specifically, two challenges arise: (i) the detection algorithm should account for two dimensional spatio-temporal temperature dynamics, and (ii) temperature sensor placement should consider detectability and isolability of the faults. Furthermore, localization of thermal faults within such large format batteries can be useful from thermal management perspective. Such fault localization in large battery cells have been under-explored. These challenges necessitate additional research on thermal diagnostics in large format pouch or prismatic cells.

\subsection{Literature review and research gaps}
In recent years, battery thermal fault diagnostics have received considerable awareness as many model-based approaches \cite{marcicki2010nonlinear,dey2017model,liu2016structural,wei2019lyapunov,dey2016model,son2019model,dey2016sensor,firoozi2021cylindrical} along with data-driven and signal processing techniques \cite{ojo2020neural,hong2017big,nordmann2018thermal} have been proposed. The review papers \cite{xiong2020research,hu2020advanced,tran2020review} provide a comprehensive list of existing approaches. In \cite{marcicki2010nonlinear}, a lumped thermal model and robust estimation algorithm is used to isolate voltage, current, thermal faults and the loss of cooling system. In \cite{dey2017model}, a one dimensional distributed parameter model along with a Partial Differential Equation (PDE)-based estimator are utilized to detect and estimate thermal fault in cylindrical cells. The work in \cite{liu2016structural} isolates the voltage, current and temperature sensor faults by considering an electrical and two state thermal model for cylindrical cell. Authors in \cite{wei2019lyapunov} proposed a Lyapunov-based diagnostic algorithm for thermal parameter fault and heat generation fault in cylindrical batteries. In \cite{dey2016model}, a two-state thermal model based nonlinear observer is employed to detect and isolate thermal resistance fault, heat generation fault and convective cooling resistance fault for cylindrical cells. In \cite{son2019model}, a two-state thermal model and a stochastic Fault Detection and Diagnostic (FDD) algorithm are used to identify thermal dynamic faults such as thermal runaway in cylindrical cells. The approach in \cite{dey2016sensor} used the sliding mode observer and lumped thermal model to detect, isolate and estimate sensor faults including temperature sensor fault in cylindrical cells. Although the aforementioned research advanced the field of battery safety, they mainly focus on cylindrical cells. Thermal fault diagnostics specific to large format pouch or prismatic batteries have been under-explored.  

There are a few works in the area of fault detection and localization in battery packs. Authors in \cite{nordmann2018thermal} monitor the wire harnesses in battery pack to obtain thermal hot-spots and localize faults based on the change in thermal characteristics of wires. However, this approach highly depends on the the types of wires and the changes in wire's physical characteristics may be influenced by other factors as well. The work in \cite{ojo2020neural} utilize the LSTM-NN based approach for thermal fault detection. In \cite{hong2017big}, a big-data and Shannon entropy analysis are used to find the time and location of faults in battery pack. However, the data-driven approaches \cite{ojo2020neural,hong2017big} suffer the drawbacks of large data requirement and limited ability to capture unforeseen anomalies. The work in \cite{xu2020internal} shows that the surface temperature of pouch cells is a proper indicator for thermal runaway detection and is an appropriate replacement of internal temperature as well. However, sensor placement on cell concerning fault detection and isolation has not been investigated. In \cite{kang2020online}, a multi-fault detection and isolation approach based on voltage and current measurement in battery packs is proposed. However, fault localization has not been investigated in \cite{kang2020online}.

\subsection{Main contribution}
The main contribution of this work lies in \textit{a thermal fault detection and localization framework for large format pouch or prismatic batteries} that addresses the aforementioned gaps and challenges. We consider a physics-based two-dimensional temperature distribution model which in turn eliminates the need for large amount of training data. Within this framework, we first formulate an optimization problem for sensor placement to maximize the detectability and isolability based on knowledge of potential thermal hot-spots and given number of temperature sensors. Such sensor placement essentially partitions the two-dimensional space into multiple zones based on the number of available sensors. Subsequently, we design a bank of diagnostic filters where each filter corresponds to each zone. Each of these filters is designed using Kalman filtering framework and the innovation sequences of these filters are used to detect and isolate faults in each zone. 

The rest of the paper is organized as follows. Section 2 presents the thermal dynamics model and Sections 3 and 4 discuss the proposed detection and isolation scheme. Section 5 presents the case studies on a commercial cell. Finally, Section 6 concludes the work. 
\section{Modeling Framework and Problem Statement}
Consider a large format battery with {length $\mathcal{M}$, breadth $\mathcal{N}$,} and depth $p$. Since $p << \mathcal{M}$ and $p<<\mathcal{N}$, we consider the temperature distribution along length and breadth, denoted by reference $x$ and $y$ axes {\cite{Yazdanpour,guo2013distributed,jaguemont2017development}}. In this setting, the two dimensional temperature distribution $\Theta(x,y,t)$ along $x$ and $y$ axes can be represented by a Partial Differential Equation (PDE) \cite{Yazdanpour}: 
\begin{align}
    &{\rho C_p}\frac{\partial \Theta}{\partial t} =  k\left[\frac{\partial^2 \Theta}{\partial x^2} + \frac{\partial^2 \Theta}{\partial y^2}\right]+\dot{q}-\dot{q}_h, \label{PDE}\\\label{q}
    &\dot{q}=g(x,y)\frac{ J_a \left(E_{OCV} -E_{term}-\Theta\Gamma\right)}{v_c},\\
    &\dot{q}_h = \frac{h_oA_s(\Theta-\Theta_{amb})}{v_c},\quad \dot{SOC} =- \frac{J_a}{Q}\label{soc}
\end{align}
with the boundary conditions
\begin{align}
       & \frac{\partial \Theta}{\partial x}\Bigg|_{x=0}  = \gamma_{x0}\Big(\Theta(0,y,t)-\Theta_{amb}\Big),\label{BC1}\\
       & \frac{\partial \Theta}{\partial x}\Bigg|_{x=\mathcal{M}}  = \gamma_{\mathcal{M}}\Big(\Theta(\mathcal{M},y,t)-\Theta_{amb}\Big), \label{BC2}\\
       & \frac{\partial \Theta}{\partial y}\Bigg|_{y=0}  = \gamma_{y0}\Big(\Theta(x,0,t)-\Theta_{amb}\Big), \label{BC3}\\
     & \frac{\partial \Theta}{\partial y}\Bigg|_{y=\mathcal{N}}  = \gamma_{\mathcal{N}}\Big(\Theta(x,\mathcal{N},t)-\Theta_{amb}\Big).\label{BC4}
\end{align}

The model parameter notations are given in Table \ref{table:param}. The presented two-dimensional thermal model is formulated based on energy balance equation considering non-uniform temperature and heat distribution along length and breadth on large format cells. The generated heat inside the cell is denoted as $\dot{q}$. The term $J_a(E_{OCV} -E_{term})$ in \eqref{q} represents the Ohmic heat generation and the third term is the entropic heat \cite{samba2014development}. The function $g(x,y)$ accounts for the non-uniform heat generation. Moreover, heat generation is neglected along the depth of the cell since $p << \mathcal{M}$ and $p<<\mathcal{N}$. The boundary conditions are based on the heat dissipated from the surface of the battery to the surrounding area which is at ambient temperature $\Theta_{amb}$. The $\dot{q}_h$ is the transverse heat flux which is obtained based on the Newton's law of cooling. In transverse heat flux modeling we consider the convection heat while neglecting the radiation heat. The second equation in \eqref{soc} is the State of Charge (SOC) dynamic obtained from Coulomb counting and Q is the battery capacity in $(A-s)$. Furthermore, we assume that the entropic heat coefficient is negligible {as its magnitude is small and the overall contribution of  entropic heat is much smaller than the other kind of heat. This assumption enables us to utilize a linear model setting which simplifies the analytical derivation of the proposed detectability and isolability conditions. Furthermore, the effect of entropic coefficient is captured in the process noise while designing the Kalman filter based detection and isolation scheme.} Finally, we follow the convention that applied current ${J_a}$ to be positive during discharge.   

\subsection{Thermal faults}
There are several phenomena that cause anomalous heat generation and consequently overheating of batteries \cite{tran2020review}. For instance, one of the possible causes can be the chemical reactions which damage the separator and cause internal short circuit leading to excessive heat. Some other causes of heat generation can be external short circuit coming from a heat path between the cathode and anode, electrolyte leakage or a load with very small resistance. The cooling system failure can be another  cause of severe overheating. Furthermore, the voltage regulator failure, cell connection failure, charging system failure, aging under extreme temperature condition, manufacturing malfunction, {cell component reactions}, and mechanical defects can cause overheating problem as well \cite{xiong2020research,hu2020advanced,tran2020review,Bandhauer2011}. In this work, we model these anomalous heat generation as additive faults to the system, irrespective of their physical cause. \textcolor{black}{We consider both internal and external thermal faults. Internal thermal faults can potentially be generated by undesirable electrochemical reactions, internal short circuits. Internal faults are represented by an additive term in \eqref{q}, that is $\dot{q}=g(x,y)\frac{ J_a \left(E_{OCV} -E_{term}-\Theta\Gamma\right)}{v_c}+F_{int}$ where $F_{int}$ represents internal anomalous heat generation. The external faults can be caused by abnormal heat from other parts of the battery pack, e.g. neighboring cells or electrical connectors. External faults are represented by an additive term in the first equation of \eqref{soc}, that is $\dot{q}_h = \frac{h_oA_s(\Theta-\Theta_{amb})}{v_c}+F_{ext}$ where $F_{ext}$ represents external faults. Both of the variables $F_{int}$ and $F_{ext}$ represent thermal anomalies outside the normal heating. Such additive models for thermal anomalies have been commonly utilized in existing literature \cite{wei2019lyapunov,dey2017model}.}

\begin{table}[h!]
\caption{Model parameters}
\centering
\begin{tabular}{ccc}
\hline
Parameter & Description & Unit  \\[0.5ex] 
\hline
$A_s$ & Cell area & $m^2$\\
{$p$} & Cell depth & $m$\\
$v_c$ & Cell volume & $ m^3$\\
{$\mathcal{M}$} & Cell length & $m$\\
{$\mathcal{N}$} & Cell breadth & $m$\\
$J_a$ & Applied current & $A$\\
{$Q$} & Battery capacity& $A-s$ \\ 
$E_{term}$ & Terminal voltage & $V$\\
$E_{OCV}$ & Open circuit voltage & $V$\\
$\Theta_{amb}$ & Ambient temperature & $K$\\

$\Gamma$ & Entropic heat coefficient & $V/K$\\
$q_h$ & Transverse heat flux & $J$\\
$q$ & Heat generation inside the cell & $J$\\
$C_{p}$ & Average   specific heat   capacity& $J kg^{-1} K^{-1}$ \\
$\gamma$ & Boundary condition coefficients &$m^{-1}$ \\
$h_{o}$ & Effective heat transfer coefficient & $Wm^{-2}K^{-1}$ \\ 
$\rho$ & Average density of the   cell components &$kg \,m^{-3}$\\
$k$ & Average thermal conductivity of the cell & $Wm^{-1}K^{-1}$ \\

 \hline
\end{tabular}
\label{table:param}
\vspace{-2mm}
\end{table}

\subsection{Control-oriented linear thermal model and problem statement}

\begin{figure}[h!]
\centering
\subfloat[]{\includegraphics[width=0.6\columnwidth]{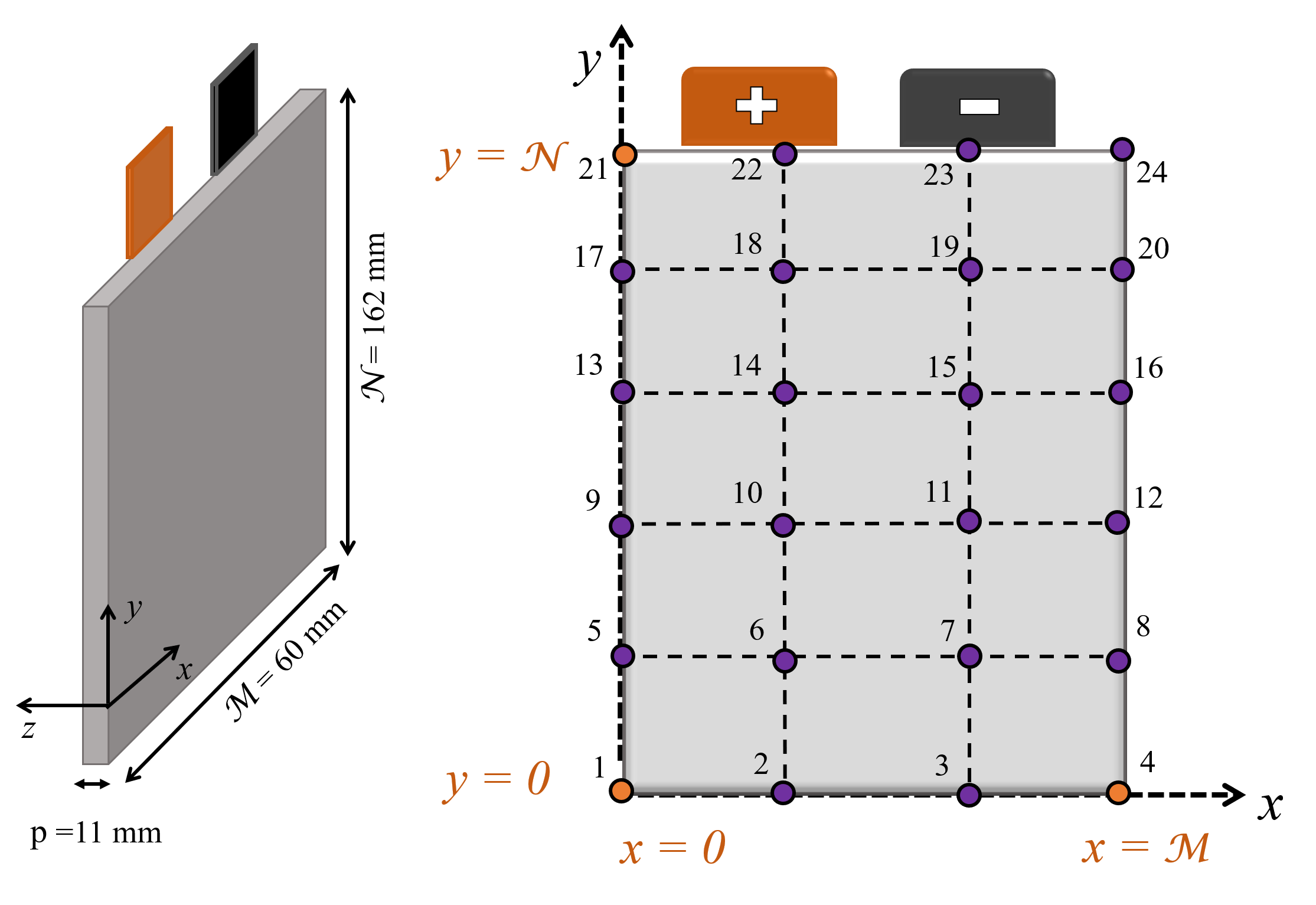}%
\label{fig:cell_coor}}
\hfil
\subfloat[]{\includegraphics[width=0.8\columnwidth]{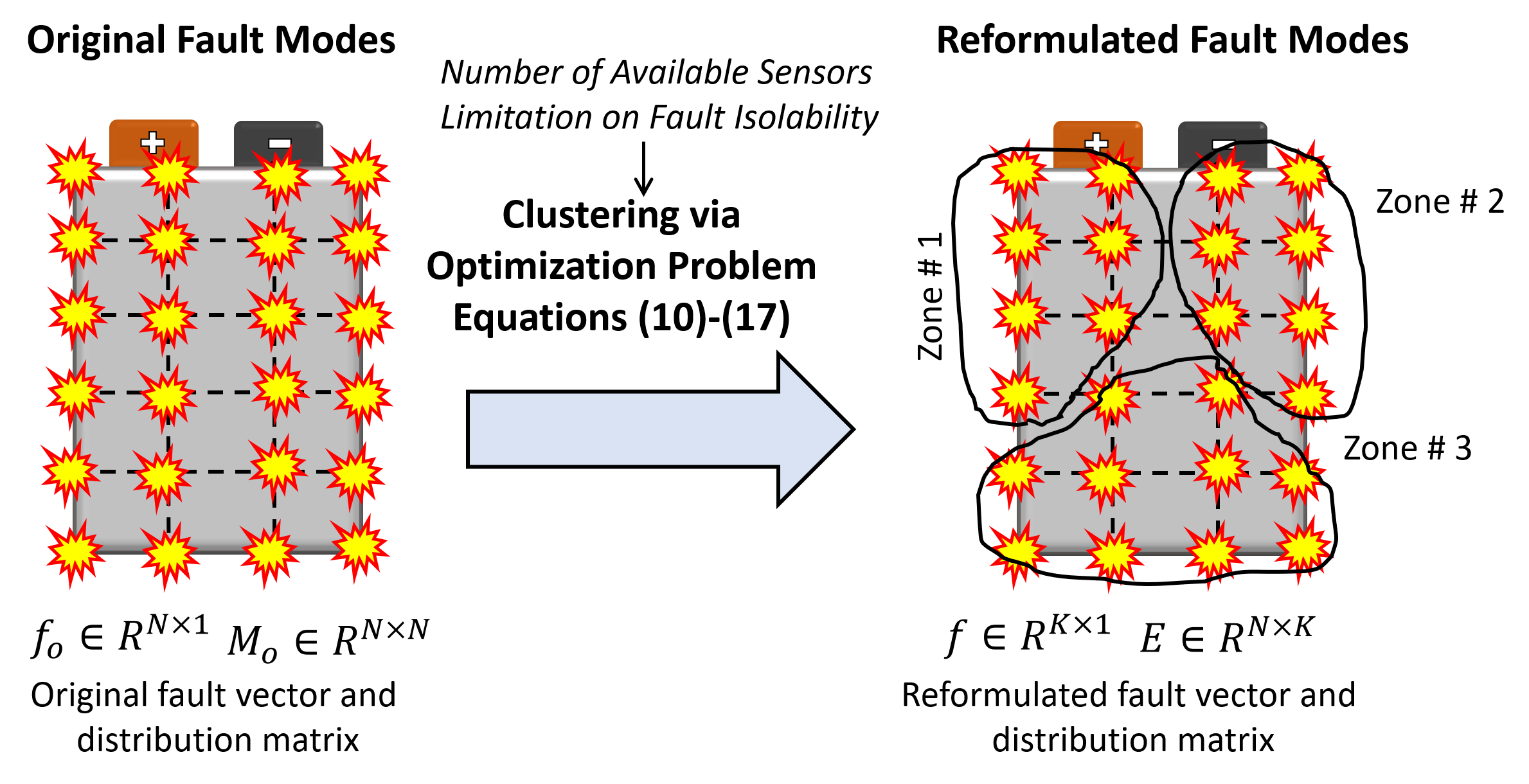}%
\label{fig:map}}
\caption{(a) Schematic of the cell under study in coordinate system with $N=24$; (b) Schematic of mapping $N=24$ fault modes to $K=3$ zones.}
\end{figure}

{The battery thermal model given in  \eqref{PDE}-\eqref{BC4}  is a PDE model which has an infinite-dimensional solution space. A common strategy to analyze such infinite-dimensional systems is to \textit{lump} the said PDE model \cite{ray_lump}. In this work, the lumping is achieved by breaking the spatial domain into sub-domains over which the state of the system is assumed to be uniform. This converts the infinite-dimensional solution space of the PDE to a finite-dimensional solution space for a set of ODEs.

Consequently, we convert the aforementioned PDE system \eqref{PDE}-\eqref{BC4} to a set of ODEs using the method of lines \cite{MoL_schiesser}. In this setting, we discretize the $x$ and $y$ spatial dimensions into a set of finite number of nodes. At each node of the system, with spatial coordinate say $(m,n)$, we lump our PDE model under the assumption of uniform temperature inside a rectangular sub-domain of size $\Delta_x\times \Delta_y$ with $(m,n)$ as center. Next, let us denote the temperature $\Theta(m,n,t)$ of a node $(m,n)$ as $\Theta_{m,n}(t)$. Thus, inside each sub-domain of the system, we can write the battery thermal model as ODEs in variables  $\Theta_{m,n}(t)$, with no spatial thermal variation.

Now under this assumption we note that the spatial thermal variation can only occur from node to node. Hence, we approximate the spatial derivatives in the PDE model into algebraic forms using the node structure.
} We approximate the first and second order spatial derivatives as $\frac{\partial \Theta_{m,n}}{\partial x} \approx  \frac{\Theta_{m+1,n}-\Theta_{m-1,n}}{2\Delta_x}$ and $\frac{\partial^2 \Theta_{m,n}}{\partial x^2} \approx \frac{\Theta_{m+1,n}-2\Theta_{m,n}+\Theta_{m-1,n}}{\Delta_x^2}$. The resulting set of ODEs represent the time evolution of temperature at the discretized nodes. {More details on PDE to ODE conversion can be found in \cite{sattarzadeh2021real}}. {The schematic of cell with discretized nodes is shown in Fig. \ref{fig:cell_coor}}.  \textcolor{black}{Subsequently, we can re-write this set of ODEs in the state-space model format \cite{palm2020system}. In systems and control analysis and design, the mathematical models are formulated in state-space format for the following advantages: (i) it provides a standard and compact representation of dynamic systems with multiple inputs and/or outputs, and (ii) it enables us to utilize standard system theoretic techniques on this format (e.g. linear algebraic tools for Ordinary Differential Equation (ODE) systems).} The resulting state-space model is given by:
\begin{align}
 \dot{T}=AT+Bu+{M}_o{f}_o, \quad y=CT, \label{ss-1}
\end{align}
\textcolor{black}{where the first equation of \eqref{ss-1} is the state dynamics equation and the second equation is the output equation. The original model \eqref{PDE}-\eqref{BC4} is reformulated into the state-space format \eqref{ss-1} in order to apply system theoretic techniques of diagnosability analysis and Kalman filter design. In \eqref{ss-1}, $T = [T_1,T_2,\cdots,T_N]^{'} \in \mathbb{R}^{N}$ is the state vector with each state $T_i$ representing the temperature of a discretized node {($T_i=\Theta_{m,n}$)} and $^{'}$ indicating transpose of a vector/matrix; $u = [\dot{q}_1,\cdots,\dot{q}_N, \Theta_{amb}]^{'}$ is the input vector which contains nominal heats $\dot{q}_1,\cdots,\dot{q}_N$ acting on nodes $1$ through $N$ with $\dot{q}_i$ being evaluated at node $i$ using \eqref{q}, and ambient temperature $\Theta_{amb}$; the variable $y = [y_1,y_2,\cdots,y_K]^{'} \in \mathbb{R}^K$ is the measurement vector with each $y_i$ representing the temperature of a node where a sensor is placed {(for example, $y_i=\Theta_{2,2}$)}. $A$ is a $N\times N$ matrix and $B$ is a $N\times (N+1)$ matrix. These matrices are derived following the approach presented in \cite{sattarzadeh2021real}. For example, with 12 ($4\times 3$) discretized nodes, these matrices are given by \cite{sattarzadeh2021real}:}
 \begin{align}
 \label{A}
    \textcolor{black}{A=\frac{k}{\rho C_p}\begin{bmatrix}\begin{smallmatrix}
    \nu_{1} & \frac{2}{\Delta_x^2} & 0&\frac{2}{\Delta_y^2}&0&0&0&0&0&0&0&0\\\frac{1}{\Delta_x^2} & \nu_{2} & \frac{1}{\Delta_x^2} &\frac{2}{\Delta_x^2}&0&0&0&0&0&0&0&0\\0& \frac{2}{\Delta_x^2}&\nu_{3}&0&0&\frac{2}{\Delta_y^2}&0&0&0&0&0&0\\\frac{1}{\Delta_y^2}&0&0&\nu_{4}&\frac{2}{\Delta_x^2}&0&\frac{1}{\Delta_y^2}&0&0&0&0&0\\0&\frac{1}{\Delta_y^2}&0&\frac{1}{\Delta_x^2}&\nu_{5}&\frac{1}{\Delta_x^2}&0&\frac{1}{\Delta_y^2}&0&0&0&0\\0&0&\frac{1}{\Delta_y^2}&0&\frac{2}{\Delta_x^2}&\nu_{6}&0&0&\frac{1}{\Delta_y^2}&0&0&0\\0&0&0&\frac{1}{\Delta_y^2}&0&0&\nu_{7}&\frac{2}{\Delta_x^2}&0&\frac{1}{\Delta_y^2}&0&0\\0&0&0&0&\frac{1}{\Delta_y^2}&0&\frac{1}{\Delta_x^2}&\nu_{8}&\frac{1}{\Delta_x^2}&0&\frac{1}{\Delta_y^2}&0\\0&0&0&0&0&\frac{1}{\Delta_y^2}&0&\frac{2}{\Delta_x^2}&\nu_{9}&0&0&\frac{1}{\Delta_y^2}\\0&0&0&0&0&0&\frac{2}{\Delta_y^2}&0&0&\nu_{10}&\frac{2}{\Delta_x^2}&0\\0&0&0&0&0&0&0&\frac{2}{\Delta_y^2}&0&\frac{1}{\Delta_x^2}&\nu_{11}&\frac{1}{\Delta_x^2}\\0&0&0&0&0&0&0&0&\frac{2}{\Delta_y^2}&0&\frac{2}{\Delta_x^2}&\nu_{12}\\
    \end{smallmatrix}\end{bmatrix}}
\end{align}
\textcolor{black}{where $\nu_{1}=-2[\frac{1}{\Delta_x^2}+\frac{1}{\Delta_y^2}+\frac{\gamma_{y0}}{\Delta_y}+\frac{\gamma_{x0}}{\Delta_x}]-A_sh_o,  \nu_{2}=-2[\frac{1}{\Delta_x^2}+\frac{1}{\Delta_y^2}+\frac{\gamma_{y0}}{\Delta_y}]-A_sh_o, 
\nu_3 = -2[\frac{1}{\Delta_x^2}+\frac{1}{\Delta_y^2}+\frac{\gamma_{y0}}{\Delta_y}-\frac{\gamma_{\mathcal{M}}}{\Delta_x}]-A_sh_o, 
\nu_4 = \nu_7 = -2[\frac{1}{\Delta_x^2}+\frac{1}{\Delta_y^2}+\frac{\gamma_{x0}}{\Delta_x}]-A_sh_o,
\nu_{5}=\nu_8=-2[\frac{1}{\Delta_x^2}+\frac{1}{\Delta_y^2}]-A_sh_o, 
\nu_6 = \nu_9 = -2[\frac{1}{\Delta_x^2}+\frac{1}{\Delta_y^2}-\frac{\gamma_{\mathcal{M}}}{\Delta_x}]-A_sh_o,
\nu_{10} = -2[\frac{1}{\Delta_x^2}+\frac{1}{\Delta_y^2}-\frac{\gamma_{\mathcal{N}}}{\Delta_y}+\frac{\gamma_{x0}}{\Delta_x}]-A_sh_o, \nu_{11} = -2[\frac{1}{\Delta_x^2}+\frac{1}{\Delta_y^2}-\frac{\gamma_{\mathcal{N}}}{\Delta_y}],
\nu_{12}=-2[\frac{1}{\Delta_x^2}+\frac{1}{\Delta_y^2}-\frac{\gamma_{\mathcal{N}}}{\Delta_y}-\frac{\gamma_{\mathcal{M}}}{\Delta_x}]-A_sh_o$, and $B=({1}/{\rho C_p})[I_N, \bar{B}]$ where $I_N$ is $N\times N$ identity matrix and $\bar{B}$ is a $N\times 1$ vector given by $\bar{B}=[2k[\frac{\gamma_{x0}}{\Delta_x} + \frac{\gamma_{y0}}{\Delta_y}] + A_sh_o,
2k[\frac{\gamma_{y0}}{\Delta_y}] + A_sh_o,
-2k[\frac{\gamma_{\mathcal{M}}}{\Delta_x} - \frac{\gamma_{y0}}{\Delta_y}] + A_sh_o,
2k[\frac{\gamma_{x0}}{\Delta_x}] + A_sh_o,
A_sh_o,
-2k[\frac{\gamma_{\mathcal{M}}}{\Delta_x}] + A_sh_o,
2k[\frac{\gamma_{x0}}{\Delta_x}] + A_sh_o,
A_sh_o,
-2k[\frac{\gamma_{\mathcal{M}}}{\Delta_x}] + A_sh_o,
2k[\frac{\gamma_{x0}}{\Delta_x} - \frac{\gamma_{\mathcal{N}}}{\Delta_y}] + A_sh_o,
-2k[\frac{\gamma_{\mathcal{N}}}{\Delta_y}] + A_sh_o,
2k[\frac{\gamma_{\mathcal{M}}}{\Delta_x} + \frac{\gamma_{\mathcal{N}}}{\Delta_y}] + A_sh_o]^{'}$. $C$ is a $K\times N$ matrix with entries $0$ or $1$ that captures the sensor locations.} The variable ${f}_o$ is the unknown fault vector affecting the node temperatures where ${M}_o$ is the fault distribution matrix. {In this work, we modelled the faults as abnormal additive heat generation. Such additive faults are represented by the vector $f_o$ which includes potential faults at each node of the pouch cell, and therefore contains N elements. Without loss of generality, we assume $f_o = [f_{o1},\cdots,f_{oN}]^{'}\in\mathbb{R}^{N\times 1}$. However, sometimes not all the nodes are equally likely to exhibit faults. For example, failure probability can be higher close to edges or cathode side under some scenarios \cite{maleki2009internal,samba2014development,mastali2018electrochemical}. Hence, we use the fault distribution matrix $M_o$ to capture this non-uniformity. The distribution matrix $M_o$ provides us a general way to capture various potential distribution of faults within the $x-y$ domain. Specific structure of $M_o$ matrix vary depending on cell chemistry and structure and can be identified utilizing battery thermal Failure Mode and Effect Analysis (FMEA) \cite{chen2019investigation,blanke2006diagnosis}} and assumed to be known in this work.
 
 {In model \eqref{ss-1}, we represent the system in state-space format which has two parts. First part is the state dynamics equation that represent the physical model. In our case, this is the temperature state dynamics originated from the two-dimensional PDE model. The second part is the output equation which captures the measured states. In our case, this is the set of nodes on the pouch cell where the thermocouples are placed. While the first part (state dynamics equation) is governed by physical behavior, the second part (output equation) is a consequence of sensor locations. In existing control literature, sensor placement is a well-known area of study. There are numerous studies on sensor placements where different objectives are considered to find the best location of sensors \cite{peddada2020optimal,chi2015sensor}. A few similar studies have also been conducted for battery applications, for instance, to improve state observability \cite{samad2015observability,wolf2012optimizing}. In this study, we consider large format battery sensor placement with the objective of improving fault detectability and isolability, which has not been explored before.} Given this setting, our goal is to develop a framework that (i) places sensors on the battery to improve fault detectability and isolability, and (ii) subsequently designs a real-time detection and isolation algorithm. Here, fault isolation indicates localization of the fault position on the battery. Non-uniform temperature distribution along with improperly placed sensors may lead to delay in detection due to fault propagation delay. Hence, it is imperative to place the sensors considering thermal hot-spots and non-uniform distribution. 

\begin{rem}[Extension to battery packs]
Although the state-space model represented in \eqref{ss-1} is derived for two dimensional large batteries, such state-space model can be extended to battery packs following the approach discussed in \cite{smyshlyaev2011pde,tian20173}. In case of a battery pack, each state would represent a node temperature in the three dimensional geometrical space. Subsequently, the proposed sensor placement and diagnostic algorithms would be applicable to such packs models with minor modifications. In this paper, as a proof of concept, we consider the two dimensional model of pouch cell as a case study to illustrate the framework.
\end{rem}

\section{Sensor placement for fault detection and isolation}
Detectability and isolability of faults are essential requirements which heavily depends on sensor locations. As discussed in Section I.A, we will discuss sensor placement for multiple sensor scenarios (relevant to safety critical applications \cite{dubaniewicz2013lithium,dubaniewicz2021thermal,faranda2019lithium, ma2021problems,caldwell2017hull}) as well as single sensor scenarios (relevant to common applications such as passenger automotive). In this work, we refer to intrinsic notion of detectability and isolability which analyze whether a fault is detectable and isolable, given a model and a set of measurements \cite{ding2008model,BulloPart1}. These metrics are defined as:

\begin{deff}[Fault Detectability]
A fault is detectable if the fault-to-output steady-state transfer gain to be non-zero. 
\end{deff}

\begin{deff}[Fault Isolability]
Two faults are isolable or distinguishable if they have distinct signatures at the output. 
\end{deff}

Such detectability and isolability conditions are determined by the potential locations of faults and locations of sensors. Here, potential fault locations indicate probable hot-spots in batteries. The placement of sensors becomes challenging when the number of available sensors is less than the number of potential fault locations. {In most of applications there are limited number of available sensors due to the practical sensor placement issues like space limitation, thickness of thermocouples, installation limitation as well as high cost and complexity of sensors \cite{chi2015sensor,rosich2007efficient}}. Additionally, the sensor placement problem is dictated by the following theoretical limitation.

\begin{limm}[Fault Isolability Limitation \cite{ding2008model}]
Given $K$ sensors, at most $K$ {additive} faults can be isolated or distinguished. 
\end{limm}

Under such theoretical limitation, we address the following question: given the knowledge of the distribution matrix $M_o$ for a probable set of $N$ hot-spots (from FMEA analysis) and a fixed number of temperature sensors $K$, how do we place the sensors on the battery such that fault detectability and isolability is ensured? 

To address this, we formulate and solve an optimization problem that (i) partitions the two dimensional battery space into $K$ zones and maps original $N$ fault modes to a new set of $K$ fault modes where each new fault mode corresponds to one particular zone and (ii) places the sensors in each zone to ensure detectability and isolability. {The partitioning requirement comes from the limited number of sensors in the cell. If we had sensors placed in each node of the cell (i.e. measure all the states), then we could have detected and isolated the faults just based on these sensor measurements. However, due to aforementioned limitations, we have limited number of sensors that can be placed on the cell. Our goal in this paper is to understand the best we can do with this limited number of sensors. Moreover, the assumption that there are $N$ hot-spots essentially means that fault can occur at any node of the pouch cell. The number $N$ comes from the fact that we have $N$ nodes and subsequently $N$ corresponding temperature state. }

In other words, we group the $N$ original fault modes into $K$ zones such that each zone has one sensor located in it. The grouping of the original faults is essential to ensure isolability {under limited number of sensors scenario} whereas placing a sensor at each zone ensures detectability. Our goal is to map the original fault vector $f_o$ with distribution matrix $M_o$ into new fault vector $f=[{{f}}_1,\cdots,{{f}}_K]^T \in{\mathbb{R}^{K}}$ with new distribution matrix $E \in \mathbb{R}^{N\times K}$. A schematic of this grouping process is shown in Fig. \ref{fig:map}. Effectively, the model \eqref{ss-1} will be modified as:
\begin{align}
& \dot{T}=AT+Bu+{E}{f}, \ y=CT. \label{ss-11}
\end{align}

Considering the reformulated model \eqref{ss-11} and previously mentioned definitions, we can write the following conditions for fault detectability and isolability \cite{ding2008model}: (i) A fault $f_i$ is detectable if  $CA^{-1}E_{i} \neq \textbf{0}$, where $E_i$ represents the $i$-th column of $E$. (ii) Two faults $f_i$ and $f_j$ are isolable if ${CA^{-1}[E_{i}~E_{j}] \neq \textbf{0}}$, where $E_i$ and $E_j$ represent the $i$-th and $j$-th columns of $E$.

To this end, our goal is to ensure fault detectability and isolability by (i) choosing the non-zero elements of $C \in \mathbb{R}^{K\times N}$ matrix, (ii) creating a new fault distribution matrix $E \in \mathbb{R}^{N\times K}$ with a new {fault vector} $f \in \mathbb{R}^{K\times 1}$ that is a reduced version of original fault vector $f_o$. {Motivated by \cite{sensor_place_fault,ding2008model},} we formulate the following optimization problem. 
\begin{align}
& \max_{C, \mathcal{S}_i} \quad \sum_{i=1}^{K} |\mathcal{G}_{i,i}|^2,\,  \label{obj}\\
   \text{ subject to: } \, & \text{rank}\begin{bmatrix}
    -A & E\\
    C & 0
    \end{bmatrix} = N+K, \label{cons-0}\\
    & C = [\mathcal{C}_{il}], \ \mathcal{C}_{il} \in  \{0,1\}, \label{cons-1}\\
    & \textstyle  \sum_l C_{il}=1,\, \forall i,\label{cons-c}\\
    & \mathcal{S}_i\subseteq \{1,\hdots, N\}, \, \cap_i \mathcal{S}_i= \{\}, \, \mathcal{S}_i\neq \{\}\label{cons-S}\\
    & E_i = \textstyle \sum_{r \in \mathcal{S}_i} {M_o}_r, \label{cons-2}\\
    & C_iE_i \neq 0, \label{cons-3}\\
   & \forall i \in \{1,\cdots,K\}, \, \forall l \in \{1,\cdots,N\}, 
\end{align}
where $\mathcal{G}=C(-A)^{-1}E$ is the fault-to-output steady-state transfer gain with $\mathcal{G}_{i,i}$ being the diagonal elements. Here, $\mathcal{C}_{il}$ are the elements of $C$ matrix; $E_i$ is the $i$-th column of $E$ matrix and $C_i$ is the $i$-th row of the $C$ matrix; and ${M_o}_r$ is the $r$-th column of $M_o$. 

{The goal of this optimization is to maximize the objective function \eqref{obj} for maximum detectability of faults. In other words, the element $\mathcal{G}_{i,i}$ which represents the static gain between fault $f_i$ and output $y_i$ is maximized here. This maximization is meaningful since higher static gains implies stronger fault signature in the output signal. Moreover, maximizing only the diagonal elements $\mathcal{G}_{i,i}$ of $\mathcal{G}$ helps fault isolation, since it forces a single fault to have a stronger effect on only one output compared to the rest. 

Subsequently, we discuss the multiple constraints imposed on this optimization problem so as to guarantee detectability and isolability of faults. The first constraint \eqref{cons-0} guarantees isolability of faults for this system \cite{ding2008model}. Among the various configurations of matrix $C$  which will guarantee system isolability in the sense of \eqref{cons-0}, we constraint the structure so that we need a single sensor to detect a single zonal fault. In order to achieve this, each row of $C$ matrix  must have a single non-zero entry. Definition of $C$ matrix in \eqref{cons-1} and constraint \eqref{cons-c} captures this structure requirement on $C$. Next, we define the non-empty set $\mathcal{S}_i$ \eqref{cons-S} which contains the indices of the columns of $M_o$ that are summed to form $i$-th column of $E$ (that is, $E_i$). Constraint \eqref{cons-2} demonstrates the strategy used to create the new fault distribution matrix $E$ from the original fault distribution matrix $M_o$ using $\mathcal{S}_i$. Effectively, each column in $E$ is created by adding several columns of $M_o$. We also note here that the construction of $E$ from $M_o$ is dictated by the isolability condition given in \eqref{cons-0}. Finally, the constraint \eqref{cons-3} ensures that the non-zero element in a row of $C$ corresponds to one of the non-zero element in each column of $E$. This essentially ensures that we have at least one output corresponds to each fault element in $f$ and detectability of that fault is guaranteed.
}
\begin{rem}[Single sensor scenario]
For the single sensor case, the measurement cannot isolate two or more faults. Hence, our goal is to obtain fault detectability \textit{only} by obtaining a \textit{single} non-zero element of vector $C \in \mathbb{R}^{1\times N}$. Moreover, we note here that the reformulated distribution matrix is the sum of all columns of the original distribution matrix i.e. $E=\sum_{r=1}^N M_{or}$ where ${M_o}_r$ is the $r$-th column of $M_o$. To this end, we could similarly define the following optimization problem:
\begin{align}
& \max_{C} \quad |\mathcal{G}|^2,\,  \\
  \text{ subject to: } &C = [\mathcal{C}_{1},\hdots \mathcal{C}_{N}], \ \mathcal{C}_{l} \in  \{0,1\}, \\
    & \textstyle  \sum_{l=1}^N C_{l}=1,
\end{align}
where $\mathcal{G}=C(-A)^{-1}E$  is a scalar fault-to-output steady-state  transfer gain.
\end{rem}

\section{Design of fault detector and isolator}
After fixing the locations of temperature sensors, we focus on the fault detector and isolator scheme. We design a bank of $K$ filters for fault detection and isolation following the idea of dedicated observer scheme \cite{clark1978instrument}. For designing the $i$-th filter, we re-write the model \eqref{ss-11} as
\begin{align}
    &\dot{T}= \begin{bmatrix}
    \dot{{T}}_i\\
    \dot{{T}}_R
    \end{bmatrix}
  =
    \begin{bmatrix}
    A_i & A_{iR}\\
    A_{Ri} & A_R
    \end{bmatrix}
    \begin{bmatrix}
    {{T}}_i\\
    {{T}}_R
    \end{bmatrix} + 
    \begin{bmatrix}
    {{B}}_i\\
    {{B}}_R
    \end{bmatrix}u   + \begin{bmatrix}
    {E}_i & 0\\
    0 & {E}_R
    \end{bmatrix}
     \begin{bmatrix}
    f_i\\
    f_R
     \end{bmatrix}, \label{ss-faulty12} \\ 
    & y  = \begin{bmatrix}
    y_i\\
    y_R
     \end{bmatrix} = 
      \begin{bmatrix}
    C_i & 0\\
    0 & C_R
     \end{bmatrix}
      \begin{bmatrix}
    T_i\\
    T_R
     \end{bmatrix} +  \begin{bmatrix}
    \vartheta_i\\
    \vartheta_R
 \end{bmatrix},\label{ss-faulty2}
 \end{align}
where $y_i$, $T_i$, and $f_i$ are the output, states, and fault located in $i$-th zone, respectively; $y_R$, $T_R$, and $f_R$ are the output, states, and faults located in rest of the zones, respectively; and $\vartheta_i$ and $\vartheta_R$ are the measurement noise corresponding to $y_i$ and $y_R$, respectively.

The condition $f_i\neq 0$ indicates there is a fault in $i$-th zone while $f_i=0$ indicates no fault. Our goal is to detect the occurrences of $f_i$ and isolate which zone is faulty. Note that one or more $f_i$ can be non-zero as multiple zones can be affected by faults. Within this bank of filters, the $i$-th filter is designed to be sensitive to $i$-th fault $f_i$ and robust to all other faults $f_R$. This condition enables both detection and isolation of the fault $f_i$. To achieve the same, we construct the $i$-th filter as
\begin{align}
     \begin{bmatrix}
    \dot{\hat{T}}_i\\
    \dot{\hat{T}}_R
    \end{bmatrix}
    = &
    \begin{bmatrix}
    A_i & A_{iR}\\
    A_{Ri} & A_R
    \end{bmatrix}
    \begin{bmatrix}
    {\hat{T}}_i\\
    {\hat{T}}_R
    \end{bmatrix}+
    \begin{bmatrix}
    {B}_i\\
    {B}_R
    \end{bmatrix}u  +
    \begin{bmatrix}
    {L}_i & 0\\
    0 & {L}_R
    \end{bmatrix}
     \begin{bmatrix}
    y_i-\hat{y}_i \\
     y_R-\hat{y}_R
    \end{bmatrix}, \label{ss-kf}\\
    \begin{bmatrix}
    \hat{y}_i\\
    \hat{y}_R
     \end{bmatrix} = 
      & \begin{bmatrix}
    C_i & 0\\
    0 & C_R
     \end{bmatrix}
      \begin{bmatrix}
    \hat{T}_i\\
    \hat{T}_R
     \end{bmatrix}, \quad    I_i = y_i-\hat{y}_i \label{ss-kf111},
\end{align}
where $I_i$  being the innovation sequence (that is, the output error) and the notation ( $\hat{}$ ) indicates filter estimate, and $L_i$ and $L_R$ are the filter gains. Subtracting \eqref{ss-kf}-\eqref{ss-kf111} from \eqref{ss-faulty12}-\eqref{ss-faulty2}, the dynamics of the innovation sequence $I_i$ can be written as:
\begin{align}
   \begin{bmatrix}
    \dot{\tilde{T}}_i\\
    \dot{\tilde{T}}_R
    \end{bmatrix}
     & =
    \begin{bmatrix}
    A_i-L_i C_i & A_{iR}\\
    A_{Ri} & A_R-L_RC_R
    \end{bmatrix}
    \begin{bmatrix}
    {\tilde{T}}_i\\
    {\tilde{T}}_R
    \end{bmatrix}
   + \begin{bmatrix}
    {E}_i & 0\\
    0 & {E}_R
    \end{bmatrix}
     \begin{bmatrix}
    f_i\\
    f_R
    \end{bmatrix}, \label{ss-kf2}\\
    I_i  & = C_i \tilde{T}_i +\vartheta_i.
\end{align}
where $\tilde{X}=X-\hat{X}, X\in \{T_i, T_R\}$. To ensure isolability, we need to have $I_i$ sensitive to $f_i$ and robust to $f_R$.

In this work, we adopt the continuous time Kalman filtering framework to design the filter \eqref{ss-kf}-\eqref{ss-kf111}. The filtering algorithm, as shown in \mbox{Algorithm \ref{algo:kalm}}, uses the following state-space model of the system:
\begin{align}
& \dot{T}={A}T+{B}{u}+G\omega, \ y=CT+\vartheta, \label{ss-k}
\end{align}
where $\omega$ is the additive disturbance with covariance $Q$ and $\vartheta$ is the measurement noise with covariance $S$. For our design, we have
\begin{align}
& A = \begin{bmatrix}
    A_i & A_{iR}\\
    A_{Ri} & A_R
    \end{bmatrix},
   \, B = \begin{bmatrix}
    {B}_i\\
    {B}_R
    \end{bmatrix},\,
   L = \begin{bmatrix}
    {L}_i & 0\\
    0 & {L}_R
    \end{bmatrix},\\
    & G = \begin{bmatrix}
    {E}_i & 0\\
    0 & {E}_R
    \end{bmatrix},\quad
    C = \begin{bmatrix}
    C_i & 0\\
    0 & C_R
     \end{bmatrix}.
\end{align}   
The covariance matrices are tuned such that $I_i$ is sensitive to $f_i$ and robust to $f_R$.

\LinesNotNumbered
\begin{algorithm}
\KwIn{$A$, $B$, $C$, $G$, $P_0$, $\bar{T}_0$, $Q$, and $S$.}
\KwOut{Filter gain $L$.}
\nextnr
Initialize error covariance matrix, $P(0) = P_0$.\\
\nextnr
Initialize states, $\hat{T}(0) = \bar{T}_0$.\\
\nextnr
Update error covariance matrix by solving $\dot{P}={A} P+ P {A}^T+ G Q G^T - PC^T S^{-1} C P$.\\
\nextnr
Compute filter gain $L = P C^T S^{-1}$.\\
\caption{{Continuous-time Kalman Filter \cite{lewis2017optimal}.}}
\label{algo:kalm}
\end{algorithm}
In summary, there will be $K$ innovation sequences available. Based on Generalized Likelihood Ratio Test \cite{willsky1976generalized}, a fault in the $i$-th zone will be detected and isolated based on the following logic: $\left|I_i\right| > \beta_i \implies \mathcal{H}_1$ and $\left|I_i\right| \leq \beta_i \implies \mathcal{H}_0$ where $\beta_i$ is a predetermined threshold, $\mathcal{H}_0$ is the null or no-fault hypothesis and $\mathcal{H}_1$ is the hypothesis indicating fault occurrence. The thresholds can be designed to satisfy prescribed False Alarm Rate (FAR) given by $R_{FA}= \mathbb{P}(f_i = 0 | \left|I_i\right| > \beta_i)$ \cite{mansouri2016statistical,dey2017model}.

\begin{rem}[Single sensor scenario]
For the single sensor case, only one filter would be designed as no isolation is possible (refer to \textit{Limitation 1} in Section III). The design of such filter would follow the same steps {given in Algorithm \ref{algo:kalm}}. However, the covariance matrices are tuned to ensure that the innovation sequence is sensitive to the faults. 
\end{rem}

\section{Case study on a commercial cell}
\begin{figure*}[h!]
\centering
\subfloat[]{\includegraphics[width=0.38\columnwidth]{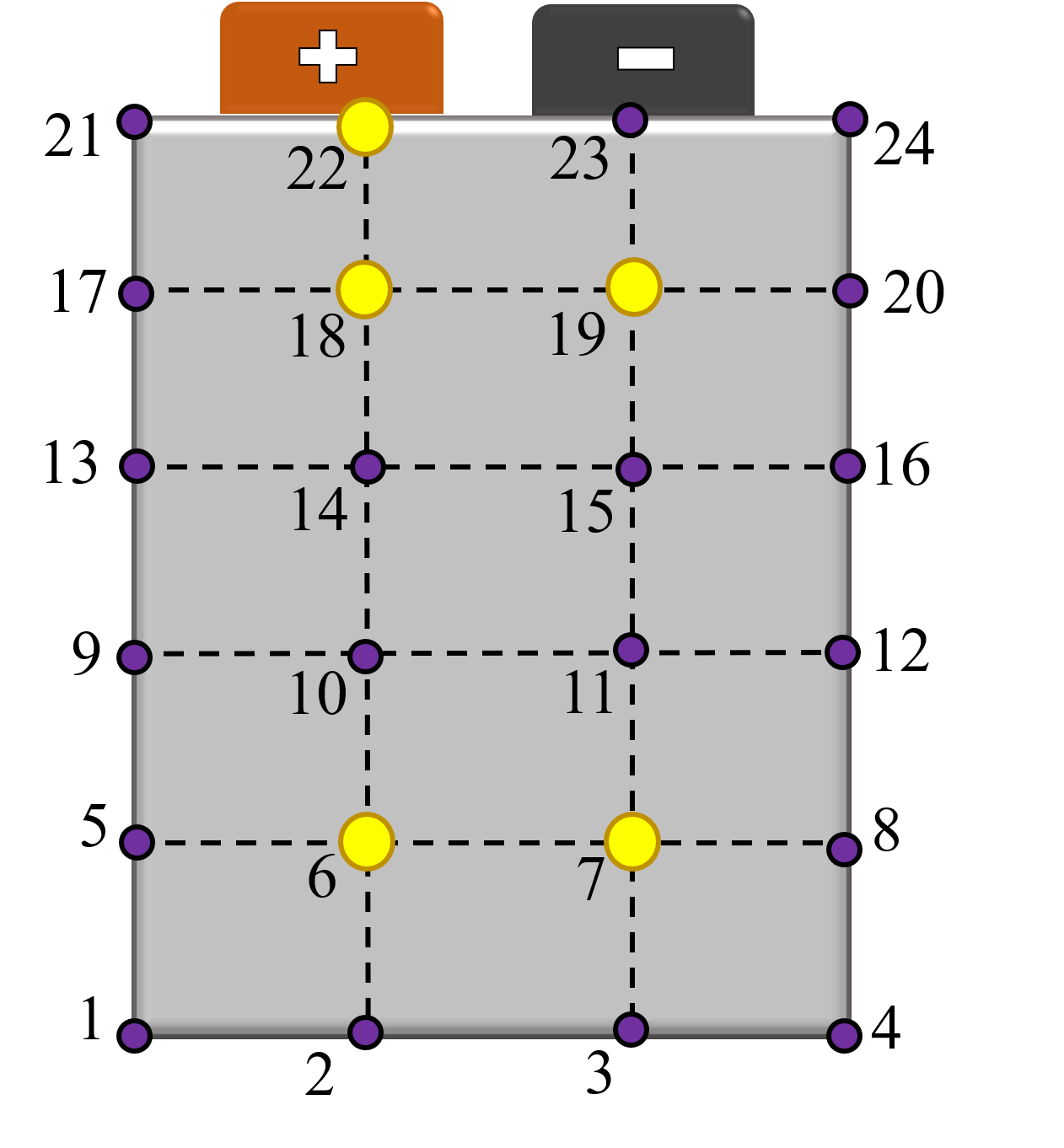}%
\label{fig:cell}}
\hfil
\subfloat[]{\includegraphics[width=0.8\columnwidth]{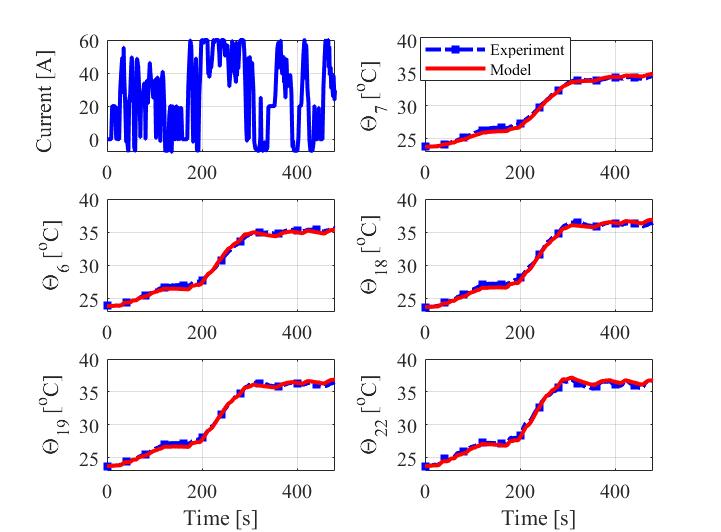}%
\label{fig:ID_result}}
\hfil
\caption{(a) Schematic diagram of the battery cell with discretized nodes. The yellow circles represent the locations of installed temperature sensors {for model identification} \cite{sattarzadeh2021real}; (b) Model identification result under no fault and modified UDDS type current profile.}
\end{figure*}

In this section, we illustrate the proposed framework via simulation and experimental studies. For this study, we adopt the pouch cell modelled and identified in \cite{sattarzadeh2021real} using Arbin BT-2000 system. The cell dimension is $60\times 162$ $mm^2$, rated capacity is $10$ Ah, and voltage range is $4.2-2.75$ $V$. We discretized the cell $x-y$ domain into $6\times 4$ nodes leading to 24 states, {\mbox{Fig. \ref{fig:cell_coor}}}, (that is, $N=24$ in \eqref{ss-1}). The cell schematic is shown in Fig. \ref{fig:cell} where the yellow circles represent the locations where temperature sensors are installed for model identification \cite{sattarzadeh2021real}. Some of the main identified parameters are: \mbox{$\gamma_n= -2.3339$ $m^{-1}$}, \mbox{$\gamma_{x0} = \gamma_{y0}=7.9746$ $m^{-1}$}, \mbox{$\gamma_{m}=-7.9746$  $m^{-1}$}, $C_{p}=1019.99$ $J kg^{-1} K^{-1}$, and \mbox{$k=5.99$ $Wm^{-1}K^{-1}$} \cite{sattarzadeh2021real}. A sample identification result for this nominal case is shown in Fig. \ref{fig:ID_result} under modified UDDS type current profile.

\begin{figure}[ht]
\centering
\subfloat[]{\includegraphics[width=0.6\columnwidth]{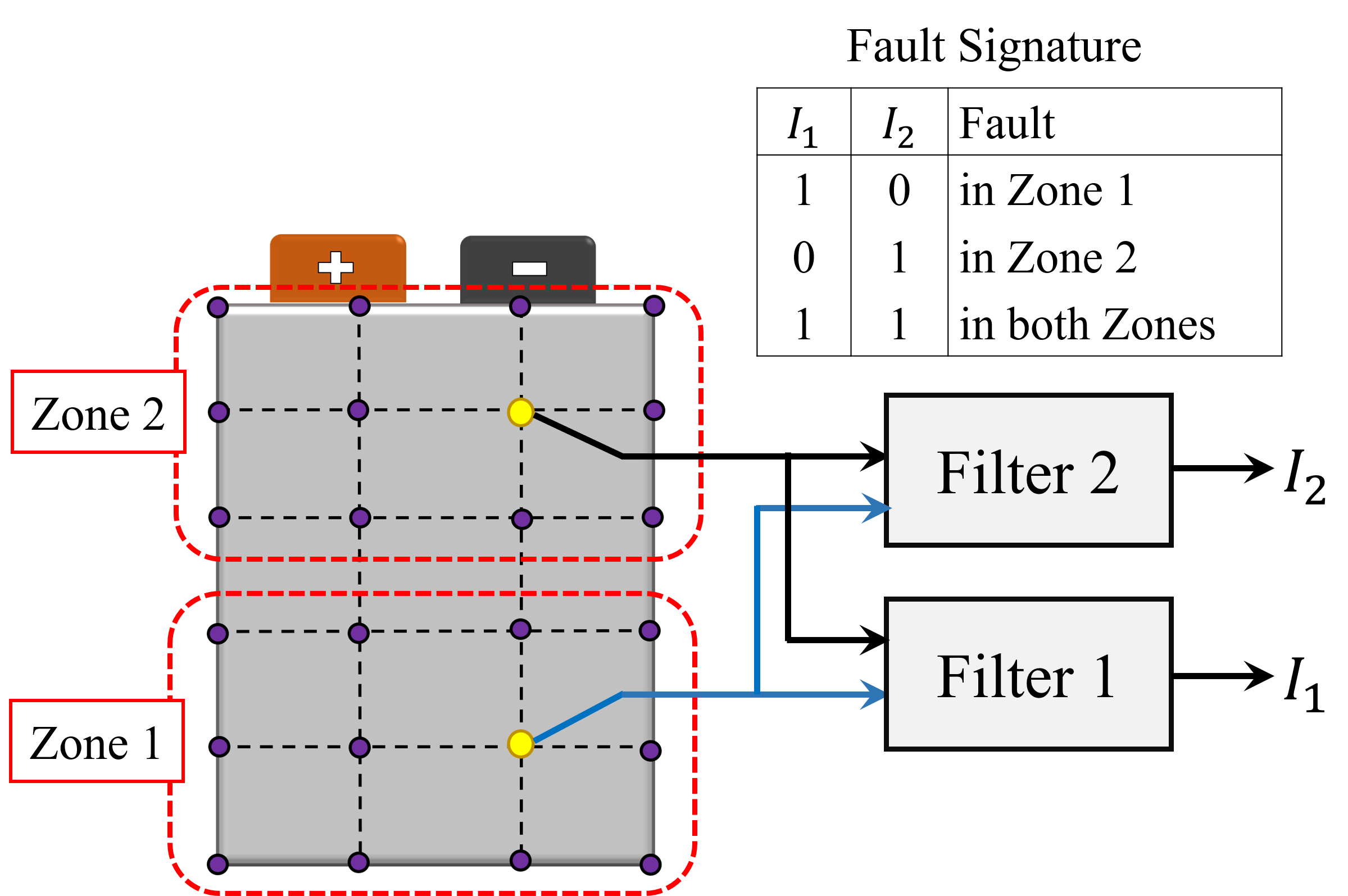}%
\label{fig:diagram}}
\hfil
\subfloat[]{\includegraphics[width=0.6\columnwidth]{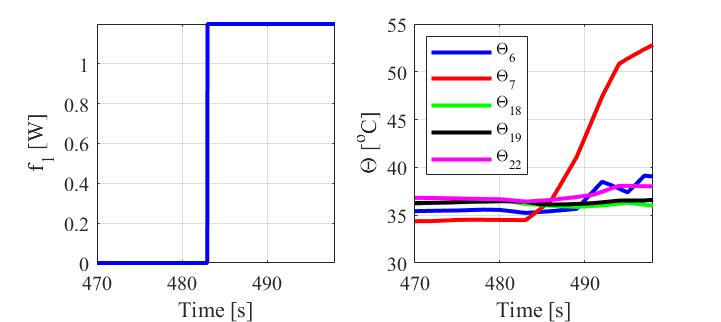}%
\label{fig:f_udds}}
\hfil
\subfloat[]{\includegraphics[width=0.6\columnwidth]{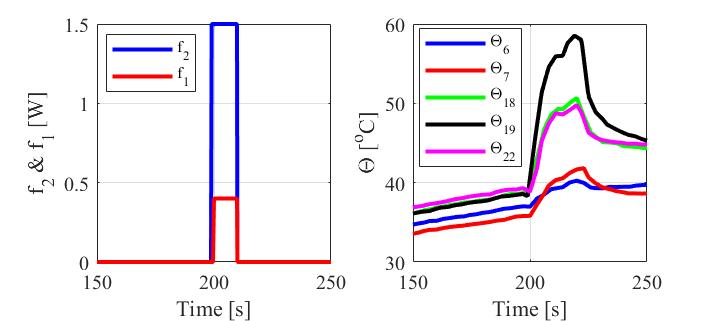}%
\label{fig:f_5c}}
\hfil
\caption{(a) Fault detection and isolation scheme; (b) Identified fault and temperature response under modified UDDS current profile. Fault injection location was Zone 1 as shown in Fig.  \ref{fig:diagram}; (c) Identified fault and temperature response under 5C constant current profile. Fault injection locations were Zones 1 and 2 as shown in Fig. \ref{fig:diagram}.}
\end{figure}

Based on this identified nominal model, we analyze the sensor placement for fault detectability and isolability as discussed in Section III.A. We assume that $M_o$ in \eqref{ss-1} is a $24\times 24$ identity matrix which indicates that fault can occur at any node of the battery. Furthermore, we assume that we can place only two sensors on the battery. Note that such choice of number of faults and number of sensors is made just to illustrate the proposed framework. This approach will still be valid for any other number of faults and sensors. Next, as discussed in Section III.A, our goal is to map these $24$ possible faults into two zones improving detectability and isolability of the faults. This analysis results in two zones with sensors placed as shown in Fig. \ref{fig:diagram}. Essentially, we have Zone 1 where node 7 is measured, and Zone 2 where node 19 is measured (see Fig. \ref{fig:cell} for node locations). Furthermore, Fig. \ref{fig:diagram} also illustrates the filter setup and the fault signature table. We design two diagnostic filters, one for each zone, and we isolate the fault based on the fault signature table in Fig. \ref{fig:diagram}. As we can see, each zone includes twelve nodes. The innovation sequence $I_1$ is sensitive to fault $f_1$ and robust to the fault $f_2$. The innovation sequence $I_2$ is sensitive to fault $f_2$ and robust to $f_1$. The filters are designed following the approach discussed in Section III.B. Based on the innovation sequence data under no fault conditions, we determined the thresholds for both $I_1$ and $I_2$ to be $\beta_1=\beta_2=0.3^oC$.

Next, we perform experiments to capture faulty scenarios. {Since it is difficult to emulate an internal fault in a safe controlled setting, we have emulated an external fault by pointing an external heater to the aforementioned zones.} {Two experiments are performed where faults are injected: first one under $5C$ constant discharge current and second one under modified UDDS type current. During each experiment, we applied the external heat by pointing the heater to certain nodes placed in the aforementioned zones} and collected the voltage, current, and temperature data. Using these data, we characterized the fault signals by minimizing the error between model data and experimental data. The characterized faults and corresponding experimental temperature data are shown in Fig. \ref{fig:f_udds} and \mbox{Fig. \ref{fig:f_5c}}. In Fig. \ref{fig:f_udds}, a constant and persistent fault was injected at $t=483$ s at Zone 1 resulting in temperature rise. In Fig. \ref{fig:f_5c}, intermittent faults were injected from $t=200$ s to $t=210$ s at both zones. In the next two subsections, we present simulation and experimental studies based on these identified models and data. 

\subsection{Simulation studies {for multiple sensors scenario}}
In simulation studies, we injected the faults in the identified simulation model and provided the faulty model data to the detector and isolator filters. We have added measurement noise of distribution $\mathcal{N}(0,\sigma^2)$, with $\sigma^2=10^{-3}$, in the model data to mimic realistic scenario. Furthermore, the filter states are initialized with incorrect initial values to verify the convergence {and the Kalman filters configured with $Q = 100 \times \mathcal{I}_{24\times24}$ at each zone, where $\mathcal{I}$ is an identity matrix. The measurement noise covariance matrix S for filter 1 is chosen to be $S_1= diag\{0.1,0.001\}$ and S matrix for filter 2 is chosen to be $S_2=diag\{0.001,0.1\}$}. We perform the following case studies to evaluate the performance of the proposed scheme under different conditions: 
\begin{itemize}
\item \textbf{Case 1:} No fault case, that is $f_1=f_2=0$.
\item \textbf{Case 2:} Intermittent pulse type fault around node 18 in Zone 2. That is, $f_1=0$ and $f_2$ is a pulse fault with magnitude $0.3$ W acting from $t=100$ s to $t=110$ s.

\item \textbf{Case 3:} Intermittent pulse type fault around node 2 in Zone 1. That is, $f_2=0$ and $f_1$ is a pulse fault with magnitude $0.7$ W acting from $t=150$ s to $t=160$ s. 
\item \textbf{Case 4:} Intermittent pulse type faults in Zone 1 around node 2 and Zone 2 around node 18. That is, $f_1$ and $f_2$ are $10$ s pulse faults of magnitude $0.5$ W and $0.3$ W, and acting at $t=150$ s and $t=100$ s, respectively.  
\item \textbf{Case 5:} Incipient type persistent faults in Zone 1 around node 2 and Zone 2 around node 18. That is, $f_1$ and $f_2$ starts at $t=100$ s and $t=150$ s, and increases slowly with time.

\item \textbf{Case 6:} No fault case under parametric uncertainty in $k$. We introduced $40\%$ uncertainty in $k$ with $f_1=f_2 = 0$.
\item \textbf{Case 7:} No fault case under parametric uncertainty in $\gamma_n$. We introduced $43\%$ uncertainty in boundary condition coefficient $\gamma_n$ with $f_1=f_2 = 0$.
\item \textbf{Case 8:} Fault in Zone 2 under parametric uncertainty in $k$. We introduced $40\%$ uncertainty in average thermal conductivity $k$ with $f_1=0$ and $f_2$ as a $10$ s pulse fault with the magnitude of $0.3$ W.
\item \textbf{Case 9:} No fault case under measurement noise of distribution $\mathcal{N}(0,\sigma^2)$ with $\sigma^2=0.01$.
\item \textbf{Case 10:} No fault case under measurement noise of distribution $\mathcal{N}(0,\sigma^2)$ with $\sigma^2=0.05$.
\item \textbf{Case 11:} No fault case under measurement noise of distribution $\mathcal{N}(0,\sigma^2)$ with $\sigma^2=0.08$.
{\item\textbf{Case 12:} No fault case under additional heat caused by other components around the cell with $F_{ext} = 4$ Watt.}
\end{itemize}

\begin{figure*}[ht]
\centering
\subfloat[]{\includegraphics[width=0.6\columnwidth]{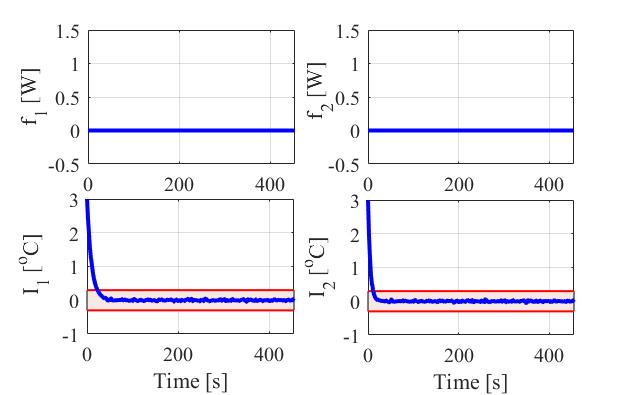}%
 \label{fig:case1}}
\hfil
\subfloat[$t=100$ s]{\includegraphics[width=0.3\columnwidth]{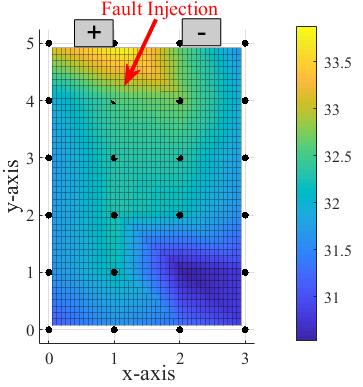}%
\label{C2_100}}
\hfil
\subfloat[$t=105$ s]{\includegraphics[width=0.3\columnwidth]{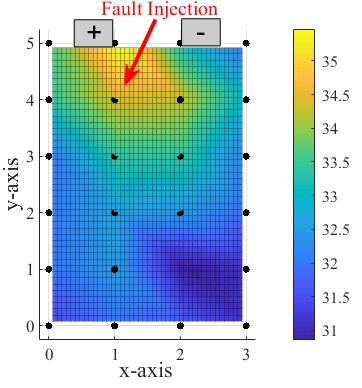}%
\label{C2_105}}
\hfil
\subfloat[]{\includegraphics[width=0.6\columnwidth]{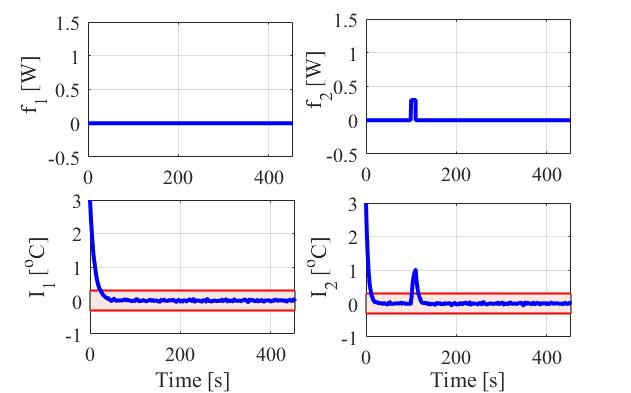}%
\label{C2}}
\caption{(a) Time evolution of fault and filter innovation sequences under Case 1; (b) Temperature distribution under Case 2 at  $t=100$ s, and (c) $t=105$ s; (d) Time evolution of fault and filter innovation sequences under Case 2 with two sensors.}
\end{figure*}

\begin{figure*}[ht]
\centering
\subfloat[$t=105$ s]{\includegraphics[width=0.3\columnwidth]{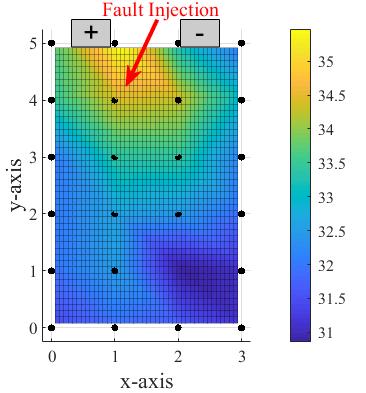}%
\label{C4_105}}
\hfil
\subfloat[$t=155$ s]{\includegraphics[width=0.3\columnwidth]{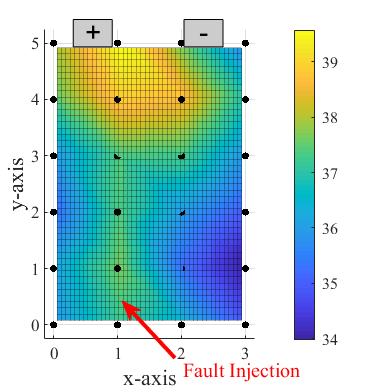}%
\label{C4_155}}
\hfil
\subfloat[]{\includegraphics[width=0.6\columnwidth]{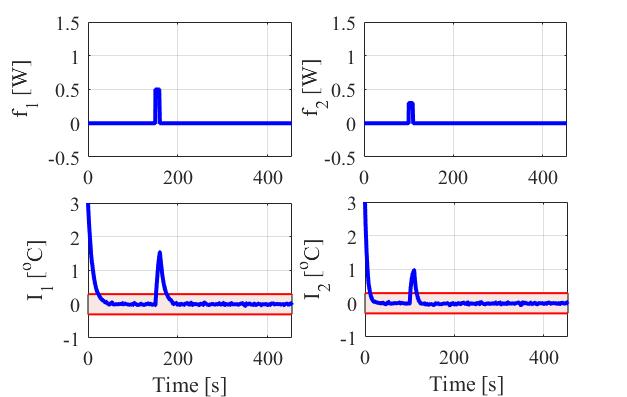}
\label{C4}}
\hfil
\subfloat[]{\includegraphics[width=0.6\columnwidth]{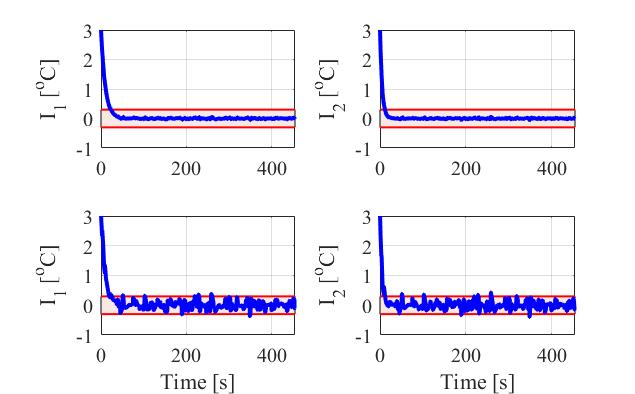}%
\label{fig:case610}}
\caption{ (a) Temperature distribution under Case 4 at $t=105$ s, and (b) $t=155$ s; (c) Time evolution of fault and filter innovation sequences under Case 4 with two sensors; (d) Time evolution of fault and filter innovation sequences under Cases 6 and 10 }
\label{Case4}
\end{figure*}

\begin{figure*}[ht]
\centering

\subfloat[$t=180$ s]{\includegraphics[width=0.3\columnwidth]{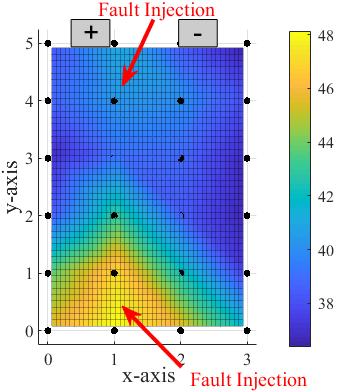}%
\label{C5_180}}
\hfil
\subfloat[$t=250$ s]{\includegraphics[width=0.3\columnwidth]{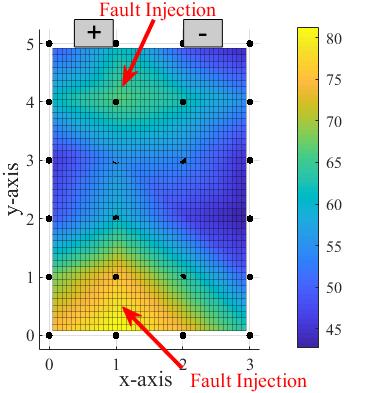}%
\label{C5_250}}
\hfil
\subfloat[]{\includegraphics[width=0.6\columnwidth]{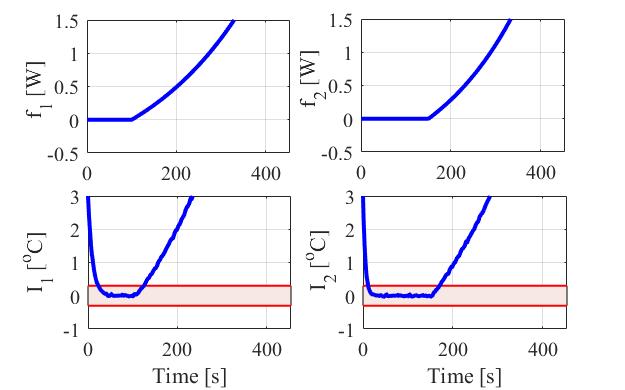}%
\label{C5}}
\hfil
\subfloat[]{\includegraphics[width=0.6\columnwidth]{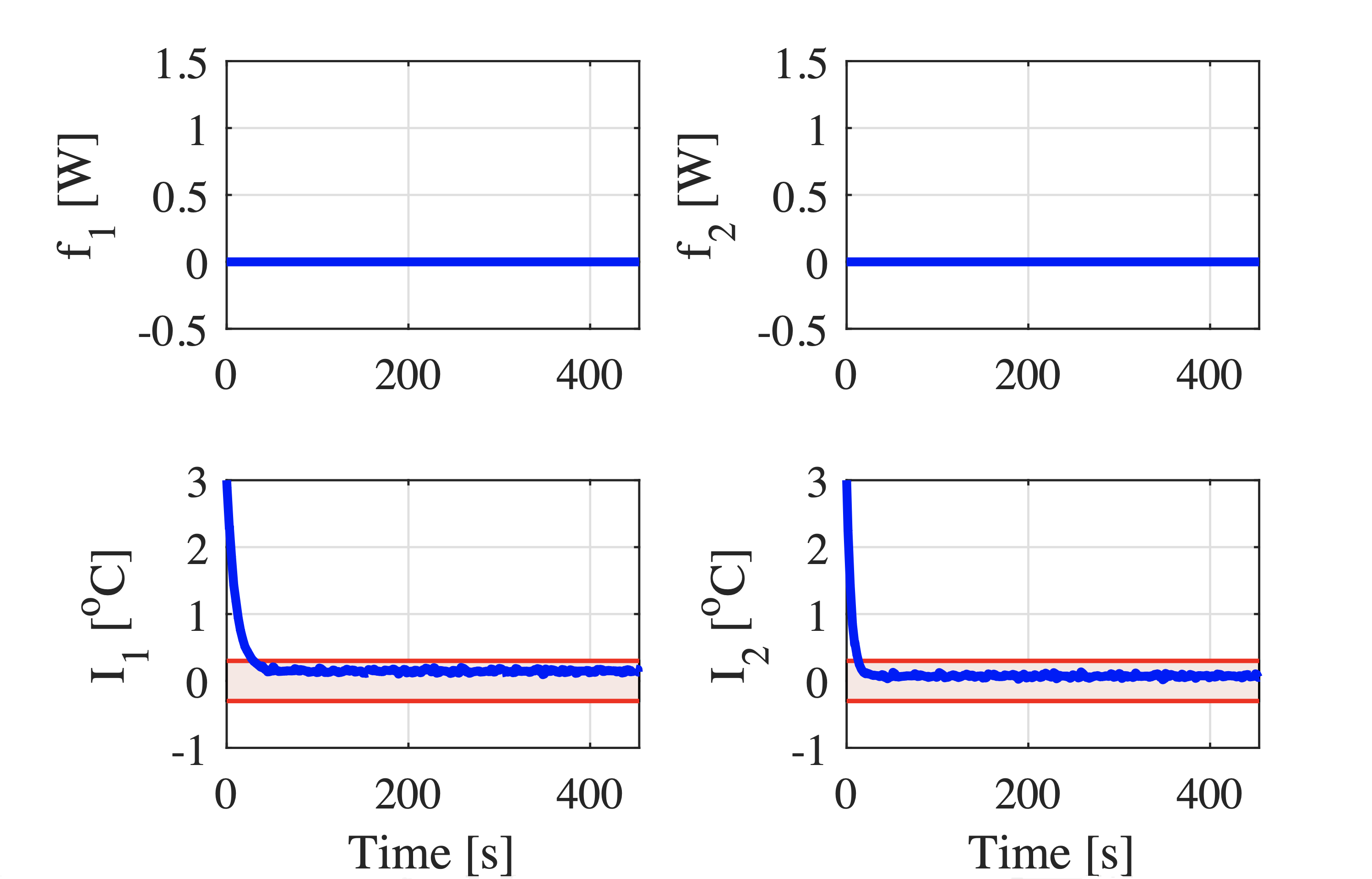}%
\label{C12}}
\caption{(a) Temperature distribution under Case 5 at $t=180$ s, and (b) $t=250$ s; (c) Time evolution of fault and filter innovation sequences with two sensors under Case 5 and; {(d) Case 12.} }
\label{Case5}
\end{figure*}

The results of these case studies are summarized in \mbox{Table \ref{table:case}} in terms of detection time, isolation performance, and false alarms. Here detection time indicates the time taken by $\left|I_i\right|$ to cross the threshold after the fault occurrence; isolation performance indicates whether the fault under consideration has been isolated; and false alarms indicate number of violations of $I_i$ under no fault conditions.  The responses of $I_1$ and $I_2$ under \textbf{Case 1} are shown in Fig. \ref{fig:case1}. As no fault is injected, $I_1$ and $I_2$ starts from non-zero initial conditions and eventually converge within the thresholds. After convergence, these signals do not cross the thresholds indicating no fault conditions. This case study verifies the convergence properties of the proposed filters. In \mbox{\textbf{Case 2}}, we have injected an intermittent fault in Zone 2 and the corresponding fault and $I_1$ and $I_2$ evolution are shown in Fig. \ref{C2_100} to Fig. \ref{C2}. As expected, $I_1$ does not cross the threshold after fault occurrence while $I_2$ crosses the threshold, detecting the fault. Furthermore, $I_2$ converges back within the threshold once the fault disappears. This case study shows that the proposed algorithm can isolate intermittent faults in Zone 2. \textbf{Case 3} shows a similar study on detection and isolation of faults in Zone 1. Next, the faults and innovation sequences under \textbf{Case 4} are shown in Fig. \ref{C4_105} to Fig. \ref{C4}. In this case, we have injected faults at both zones but at different times. The $I_1$ and $I_2$ responses indicate the both of them reacted to the fault in the corresponding zones and eventually detected them. This case study shows that the proposed algorithm performs well when there are faults in both zones. In \textbf{Case 5}, we evaluate the performance of the proposed algorithm under incipient but persistent type faults in both zones. The $I_1$ and $I_2$ responses are shown in Fig. \ref{C5_180} to Fig. \ref{C5} which confirms that both faults are detected. \textbf{Case 6} shows the performance of the proposed scheme under parametric uncertainties where the responses are shown in the top row of Fig. \ref{fig:case610}. We have added $40\%$  uncertainty in $k$ and $I_1$ and $I_2$ are shown to be robust under such uncertainty. This study shows that the proposed algorithm can handle parametric uncertainties to a certain extent. \textbf{Cases 7} and \textbf{8} study diagnostic performances with parameter uncertainties under no fault and faulty scenarios. Finally, \textbf{Cases 9-11} show the performance of the algorithm under increasing levels of measurement noise. As expected, the false alarms are increased when the noise level is higher. See Table \ref{table:case} for details. A sample of the $I_1$ and $I_2$ responses are shown in the bottom row of Fig. \ref{fig:case610} for \mbox{\textbf{Case 10}}. {In \textbf{Case 12}, we investigate the robustness of proposed algorithm with respect to the additional heat dissipating from the components around the cell such as wires. The result of this case study is shown in Fig. \ref{C12}. This study shows that the proposed algorithm is robust to heat perturbations from neighboring components and the $I_1$ and $I_2$ remain within the threshold.}

\begin{table}[ht]
\caption{Fault detection and isolation results for cases 1-11.}
\centering
\begin{tabular}{cccc}
\hline
Case No. & \makecell{False Alarms\\$I_1$, $I_2$}&  \makecell{Detection Time\\ $I_1$, $I_2$}&\makecell{Isolation\\ Performance} \\[0.5ex] 
\hline
1 & None  & N/A &N/A \\

2 & None   & N/A, 1.5 s& Isolated \\

3 & None&  1 s, N/A & Isolated  \\
4 & None & 1.4 s, 1.6 s& Isolated \\
5 & None &  25.3 s, 20.7 s& Isolated \\ 
6 & None &  N/A&N/A\\
7 & None &  N/A&N/A\\
8 & None &  N/A, 1.2 s&Isolated\\
9 & None  & N/A&N/A\\
10 & 7, 7 & N/A&N/A\\
11 & 12, 16 & N/A&N/A\\
\hline
\end{tabular}
\label{table:case}
\end{table}


\subsection{Experimental studies}
\begin{figure*}[ht]
\centering
\subfloat[]{\includegraphics[width=0.73\columnwidth]{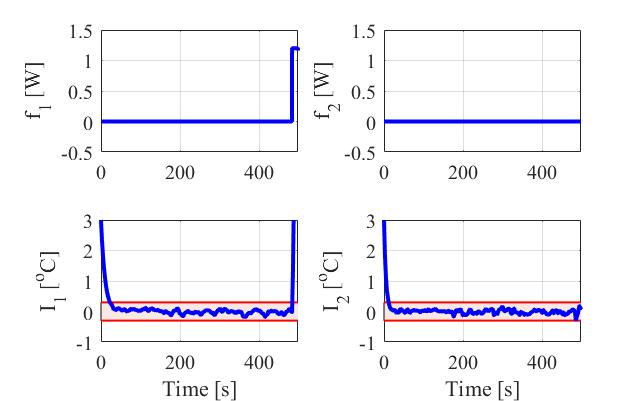}%
\label{fig:ex_result}}
\hfil
\subfloat[]{\includegraphics[width=0.7\columnwidth]{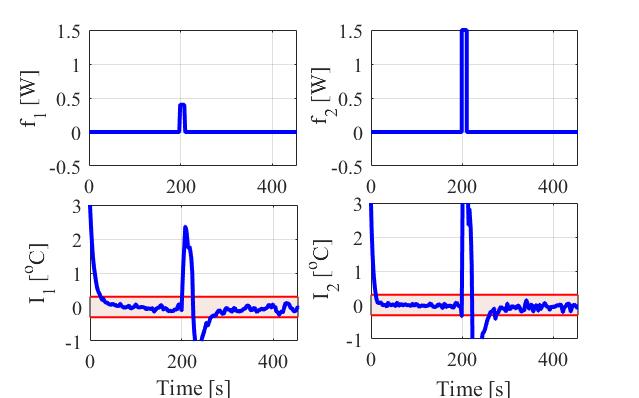}%
\label{fig:ex_result_5c}}
\hfil
\caption{(a) Experimental fault detection and isolation results under modified UDDS current profile. The fault is injected, detected, and isolated in Zone 1; (b) Experimental fault detection and isolation results under 5C discharge current profile. The fault is injected, detected, and isolated in both zones.}
\end{figure*}

In this subsection, we demonstrate the performance of proposed fault detection and localization scheme based on experimental data collected from the fault experiments discussed in the beginning of this section. We feed the measured outputs (that is, temperature data of nodes 7 and 19) to the filters. The filter states are initialized with  $3^oC$ initial error to evaluate their convergence. As discussed before, we perform two case studies. In the first case study, a persistent fault is injected at $t=483$ s at Zone 1 under modified UDDS type current (refer to Fig. \ref{fig:f_udds}). The corresponding fault and the $I_1$ and $I_2$ responses are shown in Fig. \ref{fig:ex_result}. We can see that $I_1$ and $I_2$ converge within the thresholds after starting from non-zero initial values. This shows the convergence performance of the filters under experimental data. The convergence times of $I_1$ and $I_2$ under no fault scenario are $25$ s and $13$ s, respectively. Furthermore, $I_1$ crosses the threshold after $0.5$ s of the fault occurrence and $I_2$ remains within the threshold, which shows that the proposed scheme can detect and isolate the fault in Zone 1 and the algorithm performs as intended.

In the second study, two intermittent type faults are injected at both zones from $t=200$ s to $t=210$ s under 5C constant discharge current (refer to Fig. \ref{fig:f_5c}). The corresponding faults and the $I_1$ and $I_2$ responses are shown in Fig. \ref{fig:ex_result_5c}. As it can be seen, both $I_1$ and $I_2$ cross the threshold after the fault occurrence and the faults are detected and isolated. The detection time for Zone 1 fault is $1$ s while it is $0.3$ s for Zone 2. This confirms that the proposed is able to detect and isolate multiple faults (that is, faults at both zones). Furthermore, $I_1$ and $I_2$ converge back within the thresholds in about $60$ s after the faults disappeared. This shows that the proposed scheme is able to detect and isolate intermittent type faults as well. 


\subsection{{Simulation studies for single sensor scenario}}
{In this subsection, we assume that only one sensor is available. Therefore, as mentioned in \textit{Limitation 1} in Section III, isolation is not possible. Accordingly, we solve the optimization problem discussed in \textit{Remark 2} of Section III for sensor placement that maximizes fault detectability. The analysis shows that node 14 is the best location that provides maximum fault detectability. { The filter is configured with $Q = 100 \times \mathcal{I}_{24\times24}$ and $S= 0.001$, where $\mathcal{I}$ is an identity matrix.}  Subsequently, we perform the following case studies under 5C constant current to illustrate the performance under single sensor scenario and compare the same with multiple (two) sensors scenario:
\begin{itemize}
    \item \textbf{Case I:} Intermittent  pulse  type  fault  around  node  18. That  is, $f_1= 0$ and $f_2$ is  a  pulse  fault  with  magnitude $0.3$ $W$ acting from $t= 100$ s to $t= 110s$
    \item \textbf{Case II:} Intermittent  pulse  type  fault  around  node  23. That  is, $f_1= 0$ and $f_2$ is  a  pulse  fault  with  magnitude $0.5$ $W$ acting from $t= 100$ s to $t= 110$ s
    \item \textbf{Case III:} Incipient type persistent faults around node 22. That is, $f_1$ is zero and $f_2$ starts at  $t=150$ s, and increases slowly with time.
    \item \textbf{Case IV:} Incipient type persistent faults around node 2. That is, $f_2$ is zero and $f_1$ starts at  $t=150$ s, and increases slowly with time.
    \item \textbf{Case V:} Intermittent pulse type fault around node 2. That is, $f_2=0$ and $f_1$ is a pulse fault with magnitude $0.7$ W acting from $t=150$ s to $t=160$ s.
\end{itemize}}

\begin{table}[ht]
\caption{Fault detection and isolation results for cases I-V.}
\centering
\begin{tabular}{ccc}
\hline
\makecell{Case No.\\Sensors} &   \makecell{Detection Time\\ Single, Two}&\makecell{Miss-Detection\\Single, Two} \\[0.5ex] 
\hline
I   & 2.4 s, 1.5 s &No, No \\
II &  N/A, 0.9 s & Yes, No\\

III  & 31.2 s, 27.5 s & No, No\\

IV  & 180 s, 20.9 s & No, No\\

V  & N/A, 1.1 s & Yes, No\\
\hline
\end{tabular}
\label{table:case_s}
\end{table}

\begin{figure*}[ht]
\centering
\subfloat[]{\includegraphics[width=0.7\columnwidth]{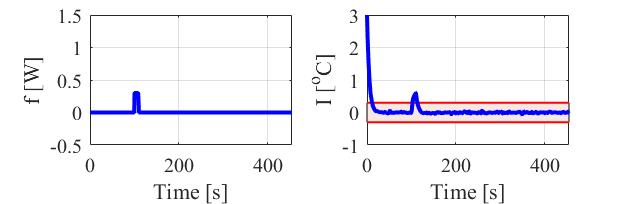}%
\label{fig:caseI}}
\hfil
\subfloat[]{\includegraphics[width=0.7\columnwidth]{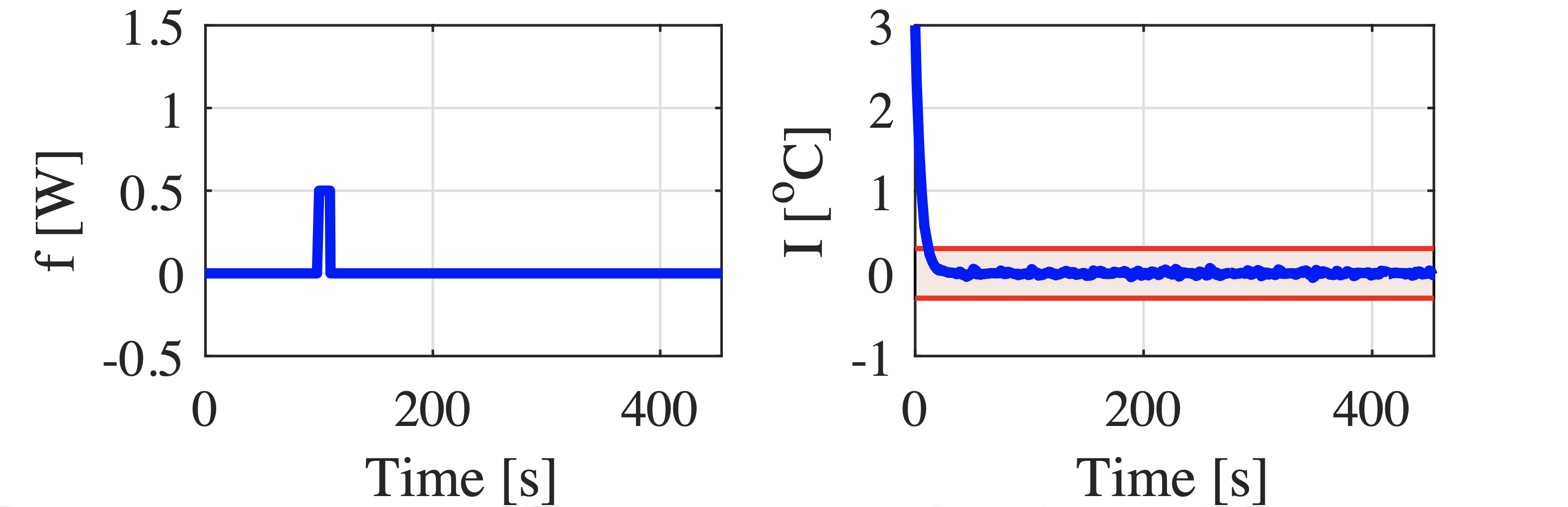}%
\label{CII_S}}
\hfil
\subfloat[]{\includegraphics[width=0.7\columnwidth]{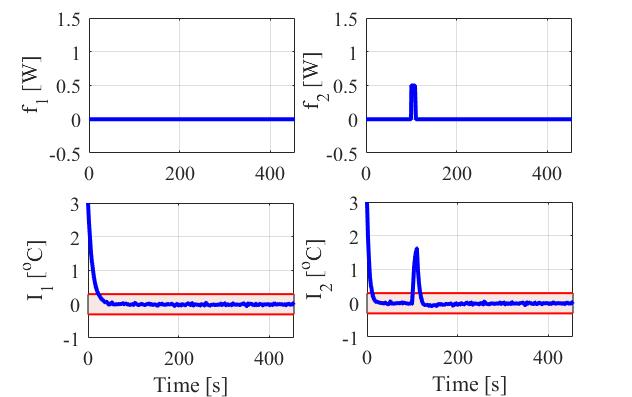}%
\label{CII_M}}
\caption{Time evolution of fault and filter innovation sequences under (a) Case I with single sensor; (b) Case II with single sensor; (c) Case II with two sensors.}
\label{CaseII}
\end{figure*}

{The results of these case studies are summarized in \mbox{Table \ref{table:case_s}} in terms of detection time and miss-detection. Here, the detection time indicates the time taken by innovation sequence to cross the threshold after fault occurrence and miss-detection indicates the scenario when the innovation sequence does not cross the threshold even in presence of a fault. The innovation sequence response under \textbf{Case I} is shown in Fig. \ref{fig:caseI}. Here, the innovation sequence crosses the threshold after 2.4 s of fault occurrence which shows the ability of fault detection using a single sensor. In \textbf{Case II}, we inject an intermittent pulse fault around node 23. The fault and innovation sequence are shown in Fig. \ref{CII_S}. As we can see, the innovation sequence with a single sensor does not cross threshold even in presence of fault at time $t=100$ s. However, with multiple sensors, the algorithm was able to detect and isolate the fault as it is shown in Fig. \ref{CII_M}. Furthermore, the detection time is slower with single sensor compared to that of the multiple sensor scenario under all the cases in Table \ref{table:case_s}. This study illustrates the fact that having multiple sensors not only enables isolability but also improves detectability and detection time. Therefore, multiple sensors are justified when safety is a critical factor compared to sensor cost.}

\section{Conclusion}
In this paper, we proposed a thermal fault detection and localization scheme for large format batteries based on a two dimensional thermal model. Within this framework, we first discussed sensor placement approach that improves fault detectability and isolability and partitions the two dimensional space into multiple zones based on given number of sensors. Thereafter, we designed a bank of diagnostic filters corresponding to each zone. The faults in each zone are detected and isolated based on the innovation sequences of these filters. We evaluated the effectiveness of proposed scheme via extensive experimental and simulation studies. Specifically, we have explored scenarios corresponding to multiple faults, intermittent faults, persistent and incipient faults, different levels of measurement noise, and different levels of parametric uncertainties. As future work, we plan to explore this framework for other types of battery faults such as electrical and electrochemical faults, and for large scale battery packs.

 \section*{Acknowledgment}
{This work was supported by National Science Foundation under Grants No. 1908560 and 2050315. The opinions, findings, and conclusions or recommendations expressed are those of the author(s) and do not necessarily reflect the views of the National Science Foundation. The authors thank the University of Colorado Denver for providing the battery testing facility. We thank Andrew Gras for his help in setting up the equipment.}


\bibliographystyle{ieeetr}

\bibliography{ref_FDI}

\end{document}